\documentclass[conference]{IEEEtran}
\IEEEoverridecommandlockouts
\usepackage{cite}
\usepackage{amsmath,amssymb,amsfonts}
\usepackage{algorithmic}
\usepackage{graphicx}
\usepackage{textcomp}
\usepackage{xcolor}
\def\BibTeX{{\rm B\kern-.05em{\sc i\kern-.025em b}\kern-.08em
    T\kern-.1667em\lower.7ex\hbox{E}\kern-.125emX}}
\begin{document}

\title{Variable Slope Trapezoidal Circulating Current Injection to Attenuate Capacitor Voltage Ripple in Modular Multilevel Converter Based Variable Speed Motor Drives Application  }

\author{\IEEEauthorblockN{Kranthi Panuganti, Rohan Sandeep Burye, and Sheron Figarado,~\IEEEmembership{Member,~IEEE}  }
\IEEEauthorblockA{School of Electrical Sciences\\ 
Indian Institute of Technology Goa\\
Goa - 403401, India\\
kranthi19242102@iitgoa.ac.in}}

\maketitle

\begin{abstract}
The main challenge in using the Modular Multilevel Converter-based constant-torque variable-speed motor drives is increased sub-module capacitor voltage ripples (SM-CVR) at low-fundamental frequency operation, due to the inverse relationship between SM-CVR and operating frequency. To address this issue, a variable slope trapezoidal circulating current (CC) is injected along with square wave common-mode voltage (CMV). Compared to sinusoidal CC and sinusoidal CMV injection, the proposed injection technique can reduce the peak of the CC in the range of 0\% to 50\%, resulting in lesser device stress and improved efficiency. Simulation results of the proposed technique are presented, and they are further compared with the existing injection techniques to show the superiority.

\end{abstract}


\section{Introduction}
In recent years, modular multilevel converters (MMCs) have been trending in medium and high power applications due to their commendable features: modularity, scalability, fault-tolerant operation, better harmonic profile [1]. These traits fascinated researchers to employ MMC for various applications. The first MMC was proposed by Marquardt et al. [2] for high voltage direct current (HVDC) transmission and later it was investigated for other applications such as medium voltage motor drives, power quality improvement, and renewable integration [1]. Also, current research is inclined towards extending these applications to electric vehicles [3], electric ships [4], and railway traction systems [5].
\\
One of the leading technical challenges in the MMC-based adjustable speed motor drives at constant-torque low-speed operation is increased magnitude of sub-module (SM) capacitor voltage ripple (CVR). This is due to the inverse dependence of SM CVR on the fundamental frequency of the MMC and this leads to increased rating values of the converter elements and affects stability [6]. Due to this, the use of MMC in variable-speed motor drives is still limited.

\begin{figure}[t!]
\centerline{\includegraphics[width=1\linewidth]{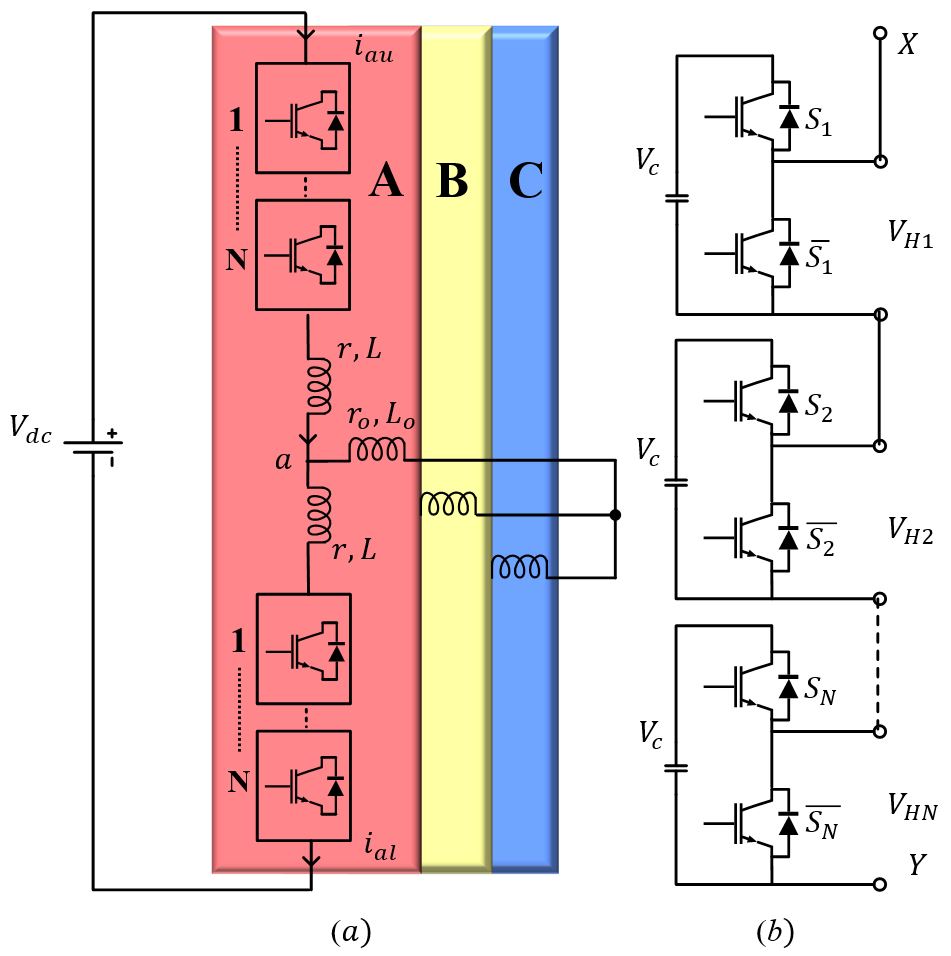}}
\caption{Configuration of MMC a) with a passive load b) series
connected SMs.}
\label{l1}
\vspace{-1em}
\end{figure}

Various researchers have proposed different techniques in the literature to address this issue over the years [7]. One of the first control techniques proposed is the injection of high-frequency common-mode voltage (CMV) and circulating current (CC) of various shapes to attenuate the CVR. An injection scheme based on sinusoidal CMV and CC was first proposed in [8].  In [8], it is shown that the peak of CC increases in MMC with respect to speed. Since injections are required mainly at low-frequency operation, tapering-off was presented for injections in [9]. In [8], injection of sinusoidal CMV and CC reflects on increased device voltage and current ratings. To address this problem, two new injection methods were proposed in [6], where method 1 used square wave CMV and sinusoidal CC and on the other hand method 2 used square wave CMV, and third harmonic injected sinusoidal CC to reduce the peak of CC to a certain level. Further, to reduce the peak of CC, a scheme was presented in [10] based on square wave CMV and CC injection. But this technique is not scalable because a square wave CC reflects in enormous magnitude of the voltage across the arm inductor at its discontinuous points, which might pose control isssues [6]. Furthermore, In [11] to reduce the peak of injected CC a relaxation on CVR is allowed instead of stringent condition to make it to zero.
\\  
This paper proposes a variable slope trapezoidal CC to reduce the peak value of CC injected.  The main advantage of proposed scheme is, using variable slope parameter $d$, CC peak value is adjusted.

\section{Modular Multilevel Converter}
The  three-phase configuration of MMC shown in Fig. 1 consists of three legs and each leg consists of two arms termed as upper arm ($u$) and lower arm ($l$) respectively. Each arm of MMC contains $N$ number of half-bridge (HB) SMs connected in series (to get $N+1$ levels in the output voltage) along with an arm inductance ($L$) which has a coil resistance ($r$). In the HB-SMs [See Fig.1b.], when $S_{1}$ is ON ($\overline{S}_{1}$ is OFF), current flows through SM capacitor, it means SM is inserted and SM voltage is $V_{Csm}$. Similarly, when $\overline{S}_{1}$ is ON ( $S_{1}$ is OFF), the current will not flow through the SM capacitor, which means that the SM is bypassed. In Fig. 2, the per-phase equivalent circuit for MMC is given and $V_{dc}$, $I_{dc}$, $v_{xs}$, and $i_{xs}$ are the DC link voltage, DC link current, output phase voltage and line current respectively. where $x$ $\in$ \{ $a$, $b$, and $c$ \} represents the phase. Similarly, the arm quantities are expressed by $i_{xy}$ and $v_{xy}$.  where $y$ $\in$ \{ $u$, $l$ \} represents an arm.
\\
Applying Kirchhoff’s voltage law to the equivalent circuit of per phase MMC in Fig. 2, and assuming that the voltage drops across $r$, $L$ are negligible, the following equations are obtained.
\begin{equation}
\begin{aligned}
V_{dc}  = v_{xu} + v_{xl} \\
v_{xs} = \frac{- v_{xu} + v_{xl} }{2}\\
\end{aligned}
\end{equation}

Similarly, applying Kirchhoff's current law expressions for circulating current ($ i_{xd} $) and output current ($ i_{xs} $) of phase $x$  are obtained and given by,
\begin{equation}
\begin{aligned}
i_{xd}  = \frac{i_{xu} + i_{xl} }{2}\\
i_{xs}  = i_{xu} + i_{xl} \\
\end{aligned}
\end{equation}

\begin{figure}[t!]
\centerline{\includegraphics[width=1\linewidth]{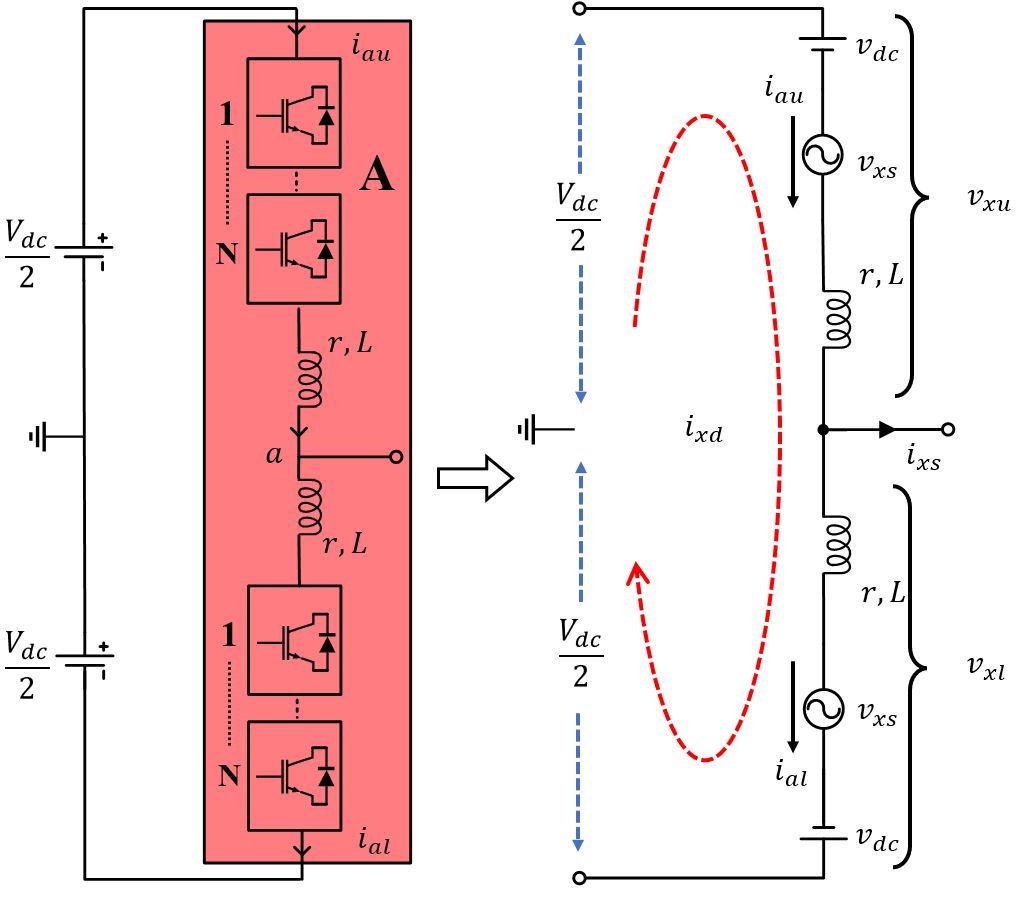}}
\caption{Per phase equivalent circuit of MMC.}
\label{l1}
\vspace{-1em}
\end{figure}
In equations 1 and 2, for sinusoidal pulse with modulation (PWM) the modulating signal, output current, and circulating currents are defined as [12]

\begin{equation}
\begin{aligned}
v_{xs}  = m_a \cdot \frac{V_{dc}}{2} \cdot \sin(\omega t)\\
i_{xs}  = I \cdot \sin(\omega t -\phi)\\
i_{xd}  = \frac{v_{xs}\cdot i_{xs}}{V_{dc}}\\
\end{aligned}
\end{equation}

Where $m_a$ is modulation index ranges from 0 to 1, $\phi$ is the initial power factor angle, $\omega$ is the system output frequency in rad/sec, and $I$ is the peak value of the output line current.

\section{Capacitor Voltage Ripple}
The arm voltage and current cause power fluctuations, leading to the SM capacitor voltage ripple. The energy stored in a single SM is given by

\begin{equation}
\begin{aligned}
E_{xy}(t)  = \int p_{xy} dt +  E_{xy}(0) \\
E_{xyh}(t) = \frac{E_{xy}(t)}{N}\\
\end{aligned}
\end{equation}
Where $E_{xy}(t)$ is energy stored in the arm, $E_{xyh}(t)$ is the energy stored in a single SM, and $h$ is the sub-module index number.  Upon substituting equation 1 to 3 in equation 4 the final expression for change in energy fluctuations given by [12],
\begin{equation}
\begin{aligned}
\Delta E_{xyh}  = \frac{1}{4Nf_s}V_{dc}I(e_{max} - e_{min})
\end{aligned}
\end{equation}
Where $e_{max}$ and $e_{min}$ are the maximum and minimum values of the function $e$, given by

\begin{equation}
\begin{aligned}
e = \frac{f^2_s - 2f^2_{sr}}{4\pi f^2_{sr}} \cos{(2\pi f_s t - \phi)} + \frac{f^2_s}{8\pi f^2_{sr}} \cos{(2\pi f_s t + \phi)} + \\
\frac{f^2_s}{24\pi f^2_{sr}} \cos{(6\pi f_s t - \phi)}
\end{aligned}
\end{equation}
where $f_s$ and $f_{sr}$ are the operating and rated frequencies. Further, change in SM energy is also given by,
\begin{equation}
\begin{aligned}
\Delta E_{xyh}  = \frac{1}{2}C_{sm}(V_c + \Delta v^x_{Cyh})^2 - \frac{1}{2}C_{sm}(V_c - \Delta v^x_{Cyh})^2
\end{aligned}
\end{equation}
By equating equations 5 and 7 , the final expression for CVR expression is derived as, 
\begin{equation}
\begin{aligned}
\Delta v^x_{Cyh}  = \frac{I}{8C_{sm}f_s}(e_{max} - e_{min})
\end{aligned}
\end{equation}
The interpretation of (8) is that CVR of SM has direct dependency on load current and inverse dependency on operating frequency. As a result under constant-torque low-fundamental frequency operation CVR is magnified.

\subsection{Injection of high-frequency common mode voltage and circulating current}

Injection of high-frequency CMV and CC is a possible solution to attenuate the CVR. It is possible to use various shapes of CMV and CC waveform injections like sine, square, and third harmonic [7].  The modified arm voltages and currents after injections are given as, 
\begin{equation}
\begin{aligned}
v_{xu}  = \frac{V_{dc}}{2} -  v_{xs} - v_{xh}  \\
v_{xl} = \frac{V_{dc}}{2} + v_{xs} + v_{xh} \\
\end{aligned}
\end{equation}
\begin{equation}
\begin{aligned}
i_{xu}  = \frac{i_{xs}}{2} +  i_{xd} + i_{xh} \\
i_{xl}  = -\frac{i_{xs}}{2} +  i_{xd} + i_{xh} \\
\end{aligned}
\end{equation}
The power fluctuations in individual SM after high frequency injections can be obtained as,
\begin{multline}
   p_{xyh}  =  \frac{1}{N}(0.5V_{dc}i_{xd} - 0.5v_{xs}i_{xs}) 
   +\frac{1}{N}(0.25V_{dc}i_{xs} - v_{xs}i_{xd}\\ \hspace{0.8cm} - v_{xh}i_{xh})
    + \frac{1}{N}(0.5V_{dc}i_{xh} - v_{xs}i_{xh} - v_{xh}i_{xd}\\ - 0.5v_{xh}i_{xs})
\end{multline}
In equation 11, term 1 is zero at the steady state condition  since the output power is equals to the input power, Also term 2 should become zero to attenuate ripple power by means of high-frequency injections. So, equating term 2 to zero yields,
\begin{table}[!t]
\caption{$k$ values for various injection techniques}
\centering
\scalebox{1.15}{%
\begin{tabular}{c c c c} 
 \hline  \hline
Technique & CMV & CC & $k$  \\
 \hline
1 & Sinusoidal & Sinusoidal & 2  \\
2 & Square & Sinusoidal & $\frac{\pi}{2} $ \\
3 & Square & Square & 1   \\
4 & Square & Proposed & 1 to 2\\
\hline \hline
\end{tabular}}
\end{table}
\begin{figure}[t!]
{\includegraphics[width=1\linewidth]{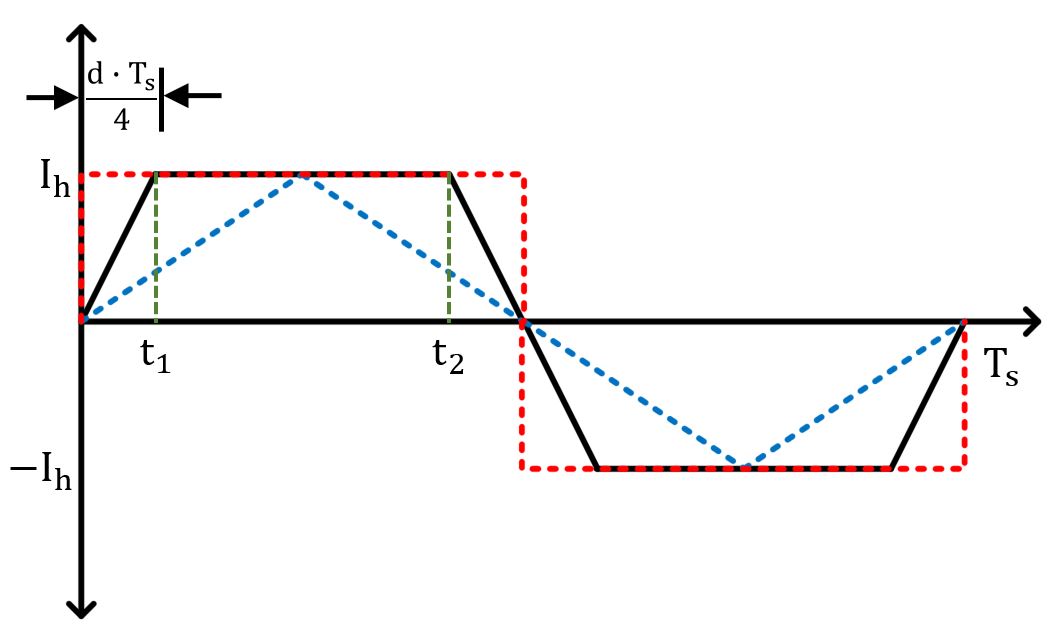}}
\caption{Variable slope trapezoidal CC waveform}
\label{l1}
\vspace{-1em}
\end{figure}

\vspace{-0.6cm}
\begin{equation}
\begin{aligned}
 v_{xh}i_{xh} = 0.25V_{dc}i_{xs} - v_{xs}i_{xd} 
\end{aligned}
\end{equation}
Further simplifying the equation 12 yields expressions for $V_h$ and $I_h$ as,
\begin{equation}
\begin{aligned}
V_h = V_{dc}(1 - m_a)\\
I_h = k\cdot k_c \cdot \frac{0.25V_{dc}i_{xs} - v_{xs}i_{xd}}{V_h}
\end{aligned}
\end{equation}
where, $V_h$ and $I_h$ are the peak values of the high-frequency CMV and CC respectively, $k_c$ is the control variable used to limit the CC injection at higher $m_a$ and $k$ is the scaling factor.
\\
In equation 13, the peak of injected CMV is chosen to depend on the operating modulation index. This is to ensure better utilization of the total voltage blocking capacity of the arm. Likewise, the peak of CC injected depends on $V_h$, and $k$. Based on the shape of injection waveform selected for CMV and CC, different values of $k$ are presented in Table I, and it is clear that the peak of $I_h$ is minimum for technique 3. However, technique 3 is not feasible in practice due to increased voltage drop across the arm inductor for square wave CC [6]. To address this problem, a variable slope trapezoidal CC technique is presented in the following subsection. Meanwhile, A detailed analysis of low-frequency operation of MMC and injection techniques is presented in [12] and [6].

\subsection{Variable slope trapezoidal CC and square wave CMV injection}
A variable slope trapezoidal CC is shown in Fig. 3, where $d$ is the variable slope parameter.  The relation between the scaling parameter $k$ and $d$ is derived in this subsection. From equation 12, the amount of average power delivered  by high-frequency components over a cycle is given by,

\begin{equation}
\begin{aligned}
P_{hf} =  \frac{4}{T_s} \int_{0}^{\frac{T_s}{4}} v_{xh} i_{xh} \cdot dt \\  \vspace{0.4cm}
 = \frac{4}{T_s} \left[ \int_{0}^{t_1} v_{xh} i_{xh} \cdot dt + \int_{t_1}^{\frac{T_s}{4}} v_{xh} i_{xh} \cdot dt \right] \\  \vspace{0.4cm}
  = \frac{4}{T_s} \left[ \int_{0}^{t_1} V_{h} \frac{I_h}{\frac{dT}{4}} t\cdot dt + \int_{t_1}^{\frac{T_s}{4}} V_{h} I_{h} \cdot dt \right]\\  \vspace{0.4cm}
   = V_h \cdot I_h \cdot (1 -0.5d)
\end{aligned}
\end{equation}
Looking at equations 13 and 14, expression for $k$ is obtained as, 
\begin{equation}
\begin{aligned}
 k = \frac{1}{(1 -0.5d)} ;\hspace{1cm} 0 \leq d \leq 1
\end{aligned}
\end{equation}

If the $d$ is zero, the trapezoid becomes the square, and $k$ has value 1. If the $d$ is 1 then trapezoid becomes the triangle, and $k$ has value $2$. 
Visualization of equation 15, which is $k$ control using the variable slope parameter $d$ is shown in Fig. 9. From Fig. 9, it is always preferred to have a lower $k$ value to have a lower peak value of CC. This proposed injection technique's advantage is that it can take the peak value of CC between technique 1 and technique 3 by varying slope parameter $d$. On the other hand, to avoid the CC discontinuity there by limiting the voltage across arm inductor, the minimum value of $d_{min}$ is limited to 0.04.

\begin{table}[!t]
\caption{MMC and Motor Parameters}
\begin{tabular}{c|c} 
\hline  \hline
 \textbf{MMC rating} & \textbf{SI Unit}\\
\hline
Output power ($P_o$) & 1 MW \\
DC link voltage ($V_{dc}$) & 7 kV\\
Number of SMs (N) & 20\\
Initial Capacitor voltage ($V_c$) & 350 V\\
Operating frequency ($f_o$) & 0-60 Hz\\
Carrier frequency ($f_{sw}$) & 500 Hz\\
Arm inductance ($L_{arm}$) & 1 mH\\
Arm resistance ($R_{arm}$) & 0.1 $\Omega$ \\
SM capacitance ($C_{sm}$) & 8 mF\\
\hline 
 \textbf{Motor Ratings} &  \textbf{SI unit}\\
\hline 
Output power ($P_s$) & 1250 hp\\
Line voltage ($V_s$) & 4160 V\\
Stator current ($I_s$) &  150 A (rms)\\
Rated speed ($N_r$) & 1189 rpm\\
Rated torque ($T_e$) & 7490 N-m\\
Number of pole pairs(p) & 3 \\ \hline \hline
\end{tabular}
\end{table}

\section{Results and Discussion}

The effectiveness of the proposed injection technique is verified through simulation, using MATLAB/Simulink, with the parameters given in Table II [12]. 

\begin{figure}[t!]
{\includegraphics[width=0.95\linewidth]{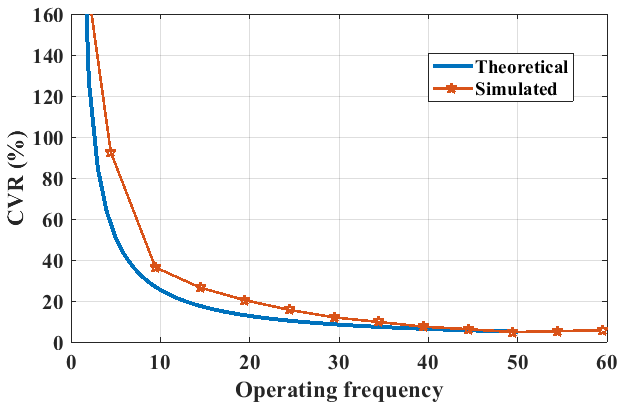}}
\caption{Variation of CVR in the percentage of SM voltage with operating frequency.}
\label{l1}
\vspace{-1em}
\end{figure}

\begin{figure}[ht!]
\centering
            \includegraphics[width=.45\textwidth]{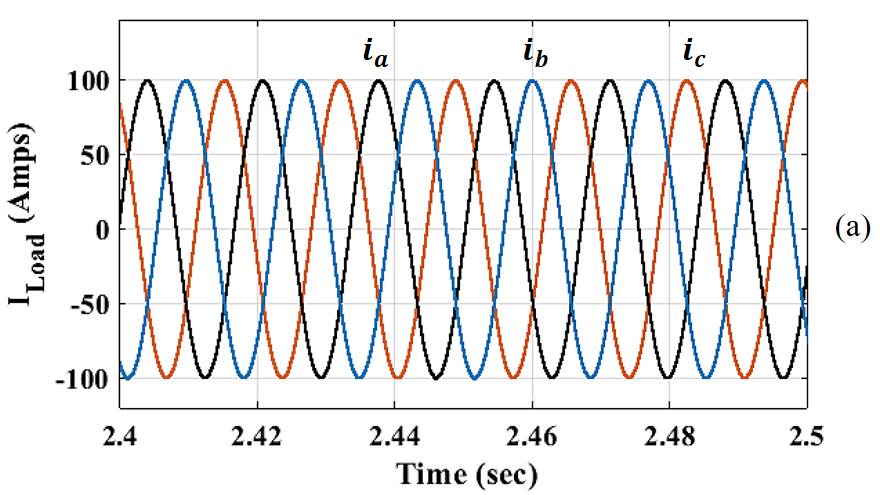}\hfill
            \includegraphics[width=.45\textwidth]{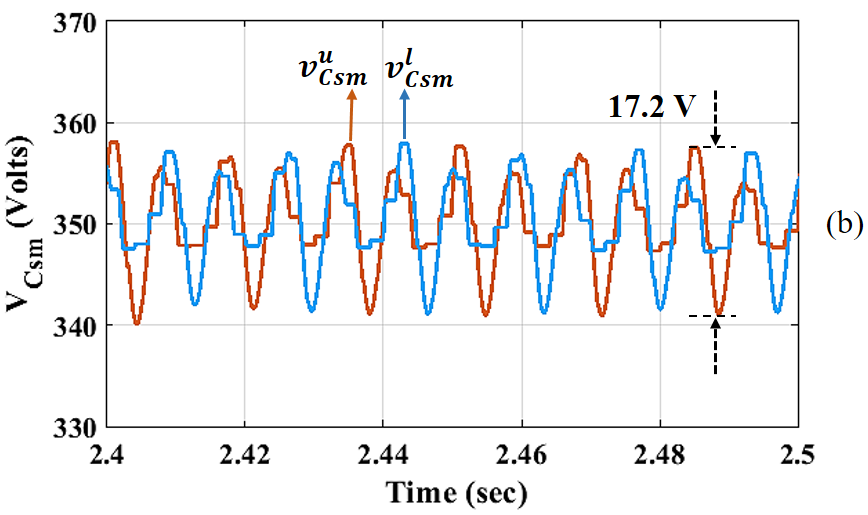}\hfill
            \includegraphics[width=.45\textwidth]{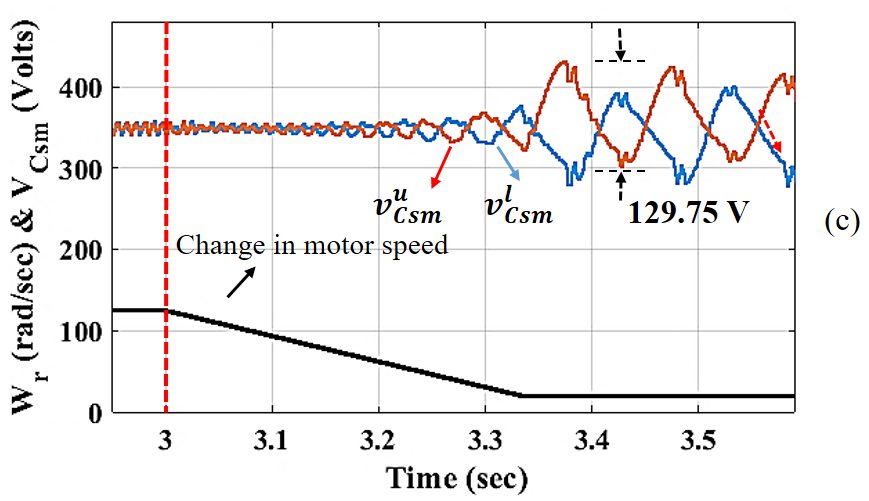}
            \caption{Output waveforms of MMC without injections a) Line currents, b) Balanced SM capacitor voltages, and c) CVR variation with speed change from rated speed to 16.6\% of rated speed, at 40\% rated load.}
\end{figure}

\begin{figure}[ht!]
\centering
            \includegraphics[width=.45\textwidth]{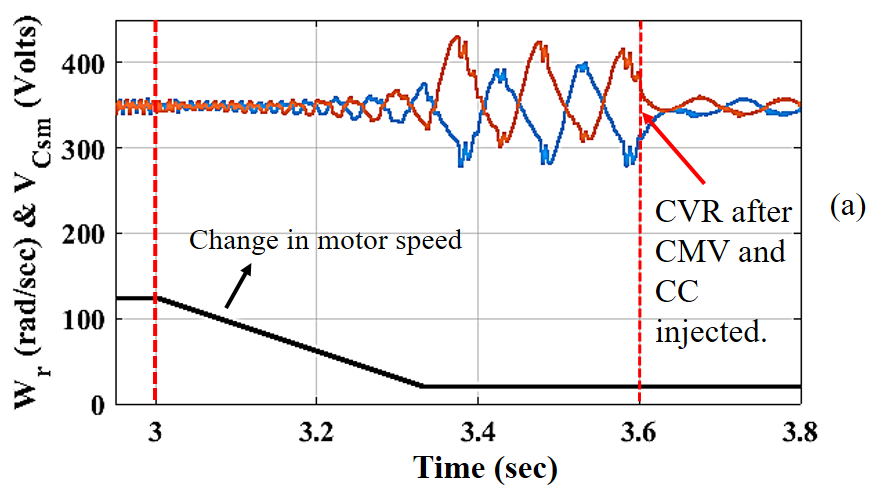}
            \includegraphics[width=.45\textwidth]{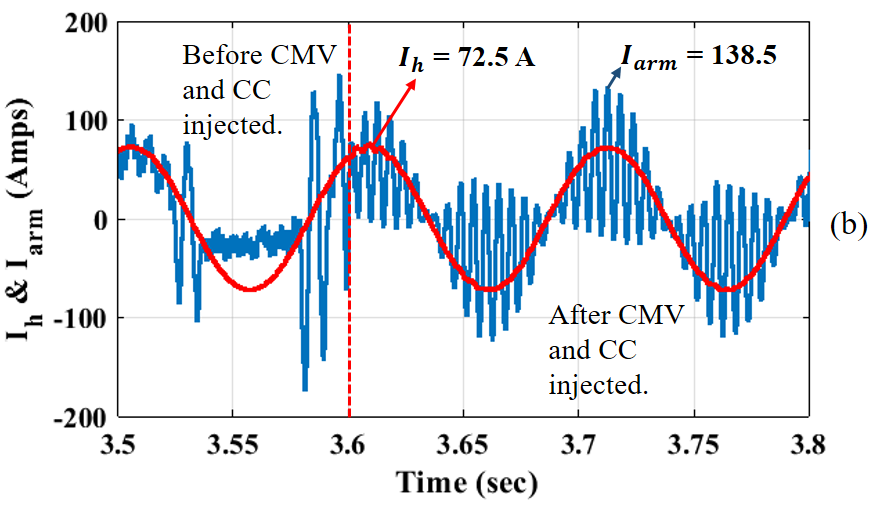}
            \caption{Output waveforms for square wave CMV and sinusoidal CC injection a) Motor speed and SM capacitor voltage, and b) Arm current and Injected CC, at 16.6\% rated speed, and 40\% rated load.}

\end{figure}
\begin{figure}[ht!]
\centering
            \includegraphics[width=.45\textwidth]{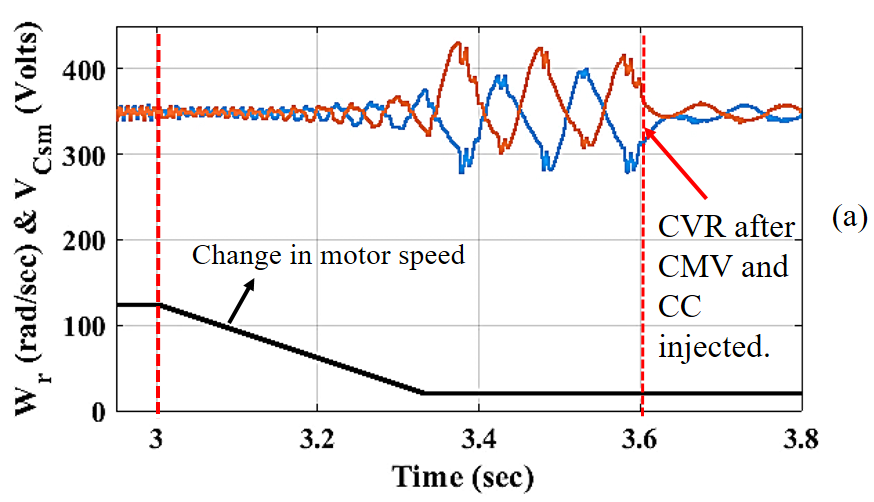}
            \includegraphics[width=.45\textwidth]{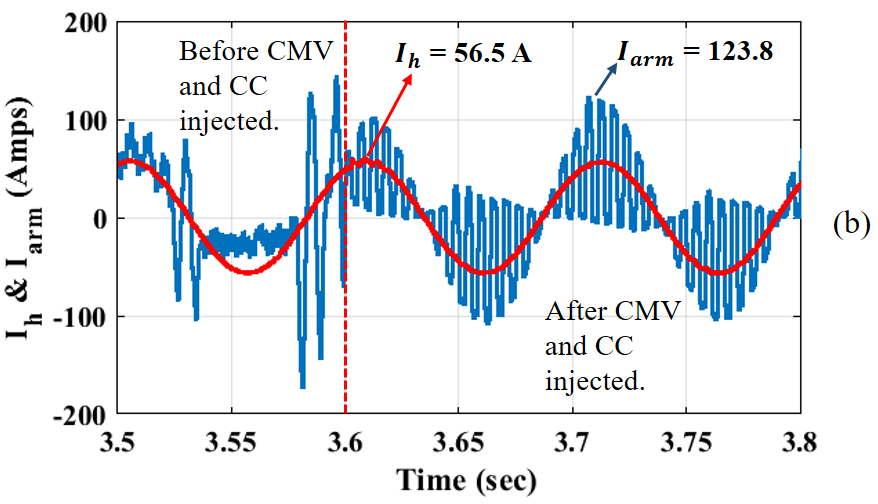}
            \caption{Output waveforms for square wave CMV and proposed CC injection a) Motor speed and SM capacitor voltage, and b) Arm current and Injected CC, at 16.6\% rated speed, and 40\% rated load.}
\end{figure}
 \begin{figure}[t!]
{\includegraphics[width=1\linewidth]{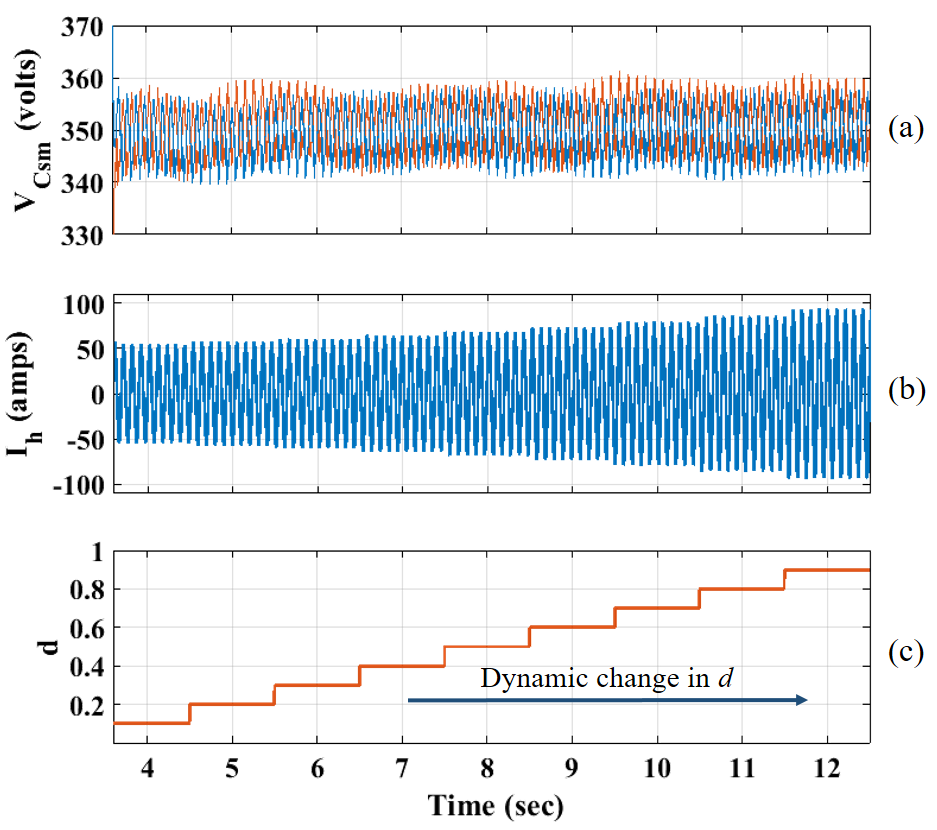}}
\caption{Converter waveforms for the dynamic change in $d$; a) CVR, b) $I_h$, and c) Change in $d$ at 16.6\% rated speed, and 40\% rated load.}
\label{l1}
\vspace{-1em}
\end{figure}

\begin{figure}[t!]
{\includegraphics[width=1\linewidth]{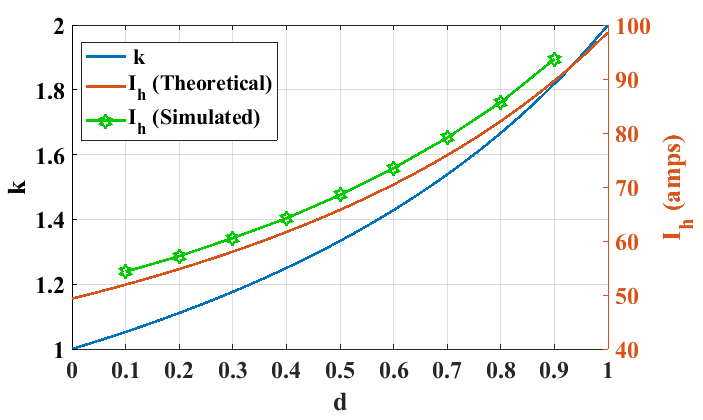}}
\caption{Variation of injected CC peak (on right) and RMS parameter (on left) with the variation in $d$.}
\label{l1}
\vspace{-1em}
\end{figure}

The MMC is connected to a 7 kV dc source and to a 1250 hp induction motor(IM) on the load side. Visualization of equation 8 and validation of it using simulations for the system in table II is shown in Fig. 4.  In Fig. 4, the variation of CVR in percentage with respect to operating frequency is shown. The curve indicates that CVR increases as the operating frequency decreases, proving the inverse dependence of CVR on operating frequency [See Eq. 8].\\
The simulated output waveforms of MMC based IM drive without injections at 100\% motor speed and 40\% rated load torque are shown in Fig. 5. Fig. 5.a shows corresponding line currents flowing into the induction motor and the peak value of line currents is 100A at 40\% of rated load torque. Fig. 5.b shows the upper arm and lower arm voltages of phase-$a$. Both the upper and lower arm SM capacitor voltages are balanced but fluctuating around 350 V with the ripple value of 17.5 V, which ensures good performance. Fig. 5.c shows CVR for both upper and lower arms for change in motor speed from 100\% to 16.6\% of the rated speed and at 40\% rated load torque. It is observed that CVR increases as the motor speed decreases and settles at 129.75 V, which is 7.4 times higher than that at the rated speed. The increased CVR affects the system reliability, and therefore high-frequency CMV and CC were injected to attenuate the increase in capacitor voltage ripple and the results are shown in Fig. 6 and Fig. 7.
\\
In Fig. 6.a, Injections were disabled intentionally during the transition of motor speed from rated speed to 16.6\% of the rated speed. On the other hand, during motor speed transition CVR increases from 17.5 V to 129.75 V. Since this is not a reliable operating point, at $t = 3.6$ seconds square wave CMV and sinusoidal CC (Technique 2) are injected to attenuate the CVR and this injection effectively attenuates the CVR to tolerable limits. This shows the effectiveness of the injected scheme. The CC and arm current waveforms for technique 2 are shown in Fig. 6.b, and the peak values of arm current and injected CC are 138.5 A and 72.5 A, respectively. Now, with the proposed injection technique, $d$ is adjusted to 0.2, and the results for the same are presented in Fig. 7. In Fig. 7.a, capacitor voltage ripple increases from 17.5 V to 129.75 V during the transition of motor speed from rated speed to 16.6\% of rated speed. at $t = 3.6$ seconds injected the square wave CMV and proposed CC which attenuates the CVR to tolerable limits same as technique 2.  However, the advantage of the proposed technique is shown in Fig. 7.b. In Fig. 7.b, arm current and injected CC has peak value of 123.8 A and 56.5 A.  In a fair comparison between proposed technique and technique 2, technique 2 injects peak CC of 72.5 A, and the proposed approach injects peak CC of 56.5 A, while both successfully attenuating the CVR to same limits. This conveys that the proposed scheme requires devices with lesser ratings to fulfill the given design constraint. 
\\
 For the proposed scheme, the dynamic variation of $d$ at 16\% rated speed and 40\% rated load is shown in Fig. 8. Here, currents settle after $d$ change in less than 0.25 seconds. It is clearly visible that as the $d$ increases in steps the peak of CC increases (i.e $I_h$) while having CVR in limits. This is consistent with Eq. 13 and Eq. 15. Furthermore, The relation between $d$, $k$ and  $d$, $I_h$ are shown in Fig. 9. In Fig. 9, simulated variation of $I_h$ validates with theoretical variation of $I_h$. The difference around 2 A in simulated and theoretical values of $I_h$ is due to simplifying assumptions made in theoretical calculations.

\section{Conclusion}
In this paper, a new injection technique is proposed to attenuate the SM CVR of MMC-based constant-torque variable-speed motor drive. The proposed strategy is based on square wave CMV and variable slope trapezoidal CC injection to reduce the SM CVR effectively. An expression relating the variable slope parameter $d$ and scaling factor $k$ is derived for the proposed technique. The major advantage of the proposed technique is that it significantly increases the efficiency by reducing the conduction losses and device current ratings of the overall drive system. The ability to achieve technique 2 and technique 3 in terms of peak of CC injected by varying $d$ is another advantage of proposed scheme. The superiority in the performance of the proposed scheme is verified through extensive simulation studies.

\end{document}